\documentclass[aps,prl,floats,twocolumn]{revtex4}
 \usepackage{graphicx}
\begin{document}
 \title{Dynamic instabilities of fracture under biaxial
strain using a phase field model} 
\author{Herv\'e Henry and Herbert.45cm
Levine}
 \date{\today} 
\affiliation{Center for Theoretical Biological
Physics\\
 University of California San Diego\\ 
9500 Gilman Dr \\
La Jolla CA, USA}
 \begin{abstract}
We  present a phase field model of the propagation of fracture under
plane strain. This model, based on simple physical considerations, is able
to accurately reproduce the different behavior of cracks (the principle of local
symmetry, the Griffith and Irwin criteria, and mode-I branching). In addition, we
test our model against recent experimental findings showing the presence of
oscillating cracks under bi-axial load. Our model again reproduces
well observed supercritical Hopf bifurcation, and is therefore the first simulation which does so.
\end{abstract}

\maketitle
 
In recent years, the physics community has seen a rebirth of interest in
the problem of dynamic fracture. This rebirth was kindled by a series of
experiments revealing that the current engineering approach to crack
propagation, namely the coupling of linear elasticity to an empirical
energy balance law for crack tip motion, cannot account for the richness
of actual fracture phenomenology\cite{marderreview}. Specifically,
dynamical instabilities which drive the system away from a single crack
propagating in a straight line require more sophisticated attention to the
actual tip region, the so-called process zone.

Given the above, it is clear that one needs a framework which can couple
local degrees of freedom involved in breaking inter-atomic bonds to global
elasticity.
 One approach, referred to as phase-field modeling of fracture,
accomplishes this task by introducing an order-parameter field (the degree
of ``broken-ness") which then couples to the elastic strain in a
manifestly continuum-level formulation. The fact that the system does not
need to be placed on a lattice avoids dynamical artifacts associated with
the breaking of translational and rotational symmetry\cite{KesslerLevine}.
One such phase-field model due to Karma, Kessler and Levine
(KKL)\cite{Karmafrac} has been shown to correctly encompass much of the
expected behavior of mode III (out-of-plane) cracks \cite{Karmabranching}.

\begin{figure}
\includegraphics[width=0.2\textwidth]{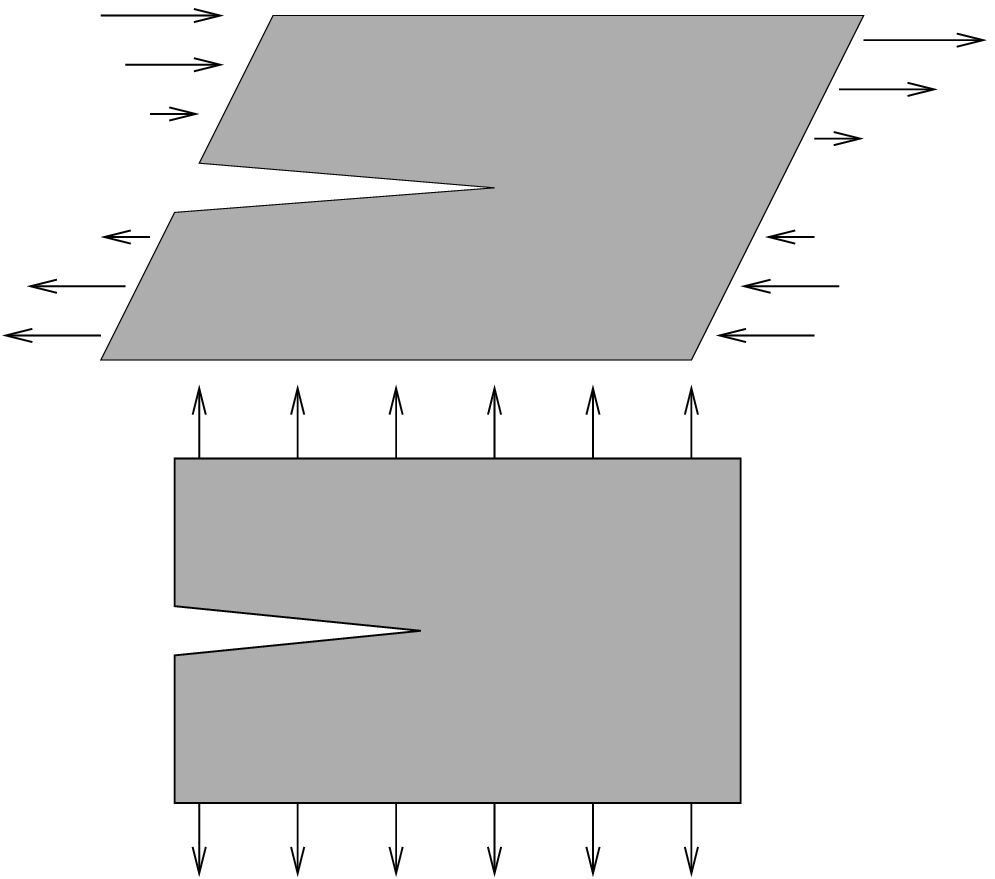}
\includegraphics[width=0.2\textwidth]{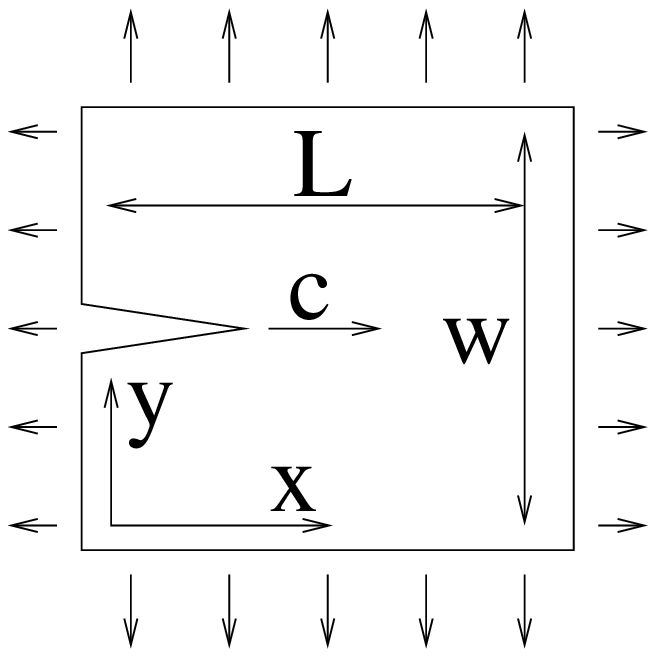}
\caption{\label{shemaswin}Left: Definition of in-plane loading modes: top
- mode II loading, bottom - mode I loading. Right: Geometry of the
experiment reported n \cite{Swinney2002}. An elastic sheet is extended in
both $x$ and $y$ directions, such that the local strain is bigger in the
$y$ direction than in the $x$ direction. The crack speed is denoted by
$c$. In  simulations performed here W was taken equal to 60 and boundary conditions at $y=0$ and $y=W$ were chosen so that  $u_x$ was kept constant in time and hence so that a vertically propagating crack can not reach $y=W$ or $y=0$.}
\vspace*{-.45cm}
\end{figure}

Here, we extend the KKL model to full vector elasticity and test its
genericity. Our major interest is in seeing whether a recently discovered
supercritical Hopf bifurcation to oscillating cracks under biaxial
loading\cite{Swinney2002} (see Fig. \ref{shemaswin}) is in fact reproduced
by KKL; we will see that in fact it is. This first-ever successful
simulation of the crack oscillation has the dual benefit of demonstrating
that the instability is not dependent on any special properties of the
specific materials used in the experiment (rubber) and also of giving us
more confidence in the phase-field methodology.

We start with the KKL phase field model\cite{Karmafrac}. Here, a sheet of
fractured elastic material is represented by the elastic displacement
field $u_x$, $u_y$ and by a phase field variable $\phi$ that can be
interpreted as the proportion of intact inter-atomic links. The evolution
equation of $u_x$ and $u_y$ derives from a modified elastic energy:
\begin{equation}
E=\int\int dx\, dy\ g(\phi)(\frac{1}{2}\lambda \epsilon_{ii}^2+\mu\epsilon_{ij}^2)\label{eq_energy}
\end{equation}
where $g=(4-3\phi)\phi^3$ is a function of $\phi$ chosen such that
$g(0)=0$, $g(1)=1$, and $g'(0)=g'(1)=0$; this specific choice is discussed
in \cite{Karmafrac}.  The tensor $\epsilon_{ij}$ is the strain tensor
which has the following form:
\begin{equation}
\epsilon_{ij}=\frac{1}{2}(\partial_i u_j+\partial_j u_i)
\end{equation}
The evolution equations for $u_x$ and $u_y$ are then:
\begin{equation}
\rho \partial_{tt}u_i=-\frac{\delta E}{\delta u_i}\label{evux}
\end{equation}
The corresponding evolution equation of $\phi$ is
\begin{equation}
\tau \partial_t \phi=\Delta \phi -\frac{d V(\phi)}{d\phi}-\frac{d g(\phi)}{d\phi}(E_\phi-\epsilon_c)
\end{equation}
where $V(\phi)=4(\phi(1-\phi))^2$ is a double well potential and
\begin{equation}
\left\{
\begin{array}{llll}
E_\phi&=& \frac{1}{2}\lambda \epsilon_{ii}^2+\mu\epsilon_{ij}^2 &\mbox{if $tr(\epsilon)>0$}\\
E_\phi&=&\frac{1}{2}\lambda \epsilon_{ii}^2+\mu\epsilon_{ij}^2-\alpha K_{lame}\epsilon_{ii}^2 &\mbox{if $tr(\epsilon)<0$}
\end{array}
\right.
\end{equation}
where $K_{Lame}=(\lambda+\mu)/2$ is the modulus of compression for a plane
strain configuration; $\alpha$ is an arbitrary coefficient chosen to be
bigger than $1$. This breaking of the symmetry between compression and
extension is a key ingredient not present in previous phase field
approaches to in-plane fracture\cite{igorphasefield,sethnaphasefield}. In
our model, If a material is simply compressed, having $\alpha>1$ will
guarantee that it will not break; also, compression will increase the
threshold needed for inducing a fracture through shear. The parameters
values used here are: $\epsilon_c=1$, $\tau=5$, $\lambda=\mu=1$ and
$\alpha=1.5$.

We proceed to study this model computationally in a box of width $W$ and
length $L$. Simulations were performed using a grid spacing of $\delta
x=0.15$ and a time step of $\delta t=0.001$ ( Decreasing to $\delta
x=0.075$ and $\delta t=0.0005$ leads to no significant difference in the
results). The time stepping scheme was the forward Euler method while the
spatial operators were computed using a discretization that conserves the
discretized energy of Eq. \ref{eq_energy} (The use of other scheme lead to long term numerical instabilities). In our simulations, we used
both fixed grids of different sizes along the $x$ axis and a grid moving
with the fracture tip along the $x$ axis. Boundary conditions for the
fixed grid were as follows: at the $y=W$ ($y=0$) boundary, $u_y$ was kept
equal to $\Delta_y$ (0) and on both lateral edges $u_x=x \Delta_x/L$. At
the $x=L$ ($x=0$) boundary, $u_x$ was kept equal to $\Delta_x$ (0) and no
flux boundaries were used for the $u_y$ field. In the case of the moving
grid, the boundary conditions at $y=W$ and $y=0$ were unchanged whereas
the boundary conditions at the horizontal ends of the grid were modified.
First we introduced an artificial viscosity in a thin layer at both ends
(This mimics partially absorbing boundary conditions). Also, at the
leftmost end the $u_x$ field was kept constant between each displacement
of the grid. We checked that those modifications did not affect the
behavior of the crack when compared to results obtained using a long
enough fixed grid, thereby allowing us to simulate the crack propagation
along an infinite strip along the $x$ axis. The initial conditions were
constructed as follows: a small initial crack was created by setting
$\phi=0$ in a small region of fixed width and variable length and by
letting the system evolve following a damped version (see later)  of the
evolution equation of elasticity (while $\phi$ was kept constant) until a
stationary state was reached.

\begin{figure}
\begin{center}
\includegraphics[width=0.23\textwidth]{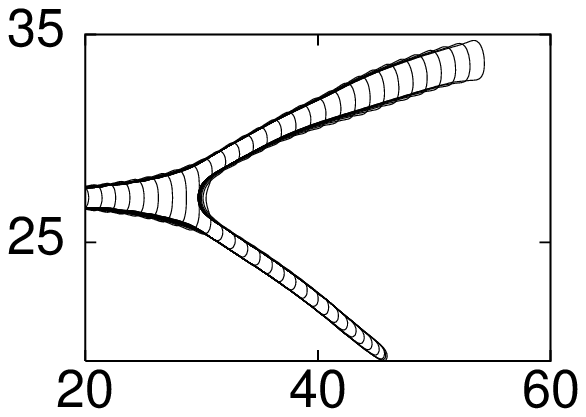}
\includegraphics[width=0.23\textwidth]{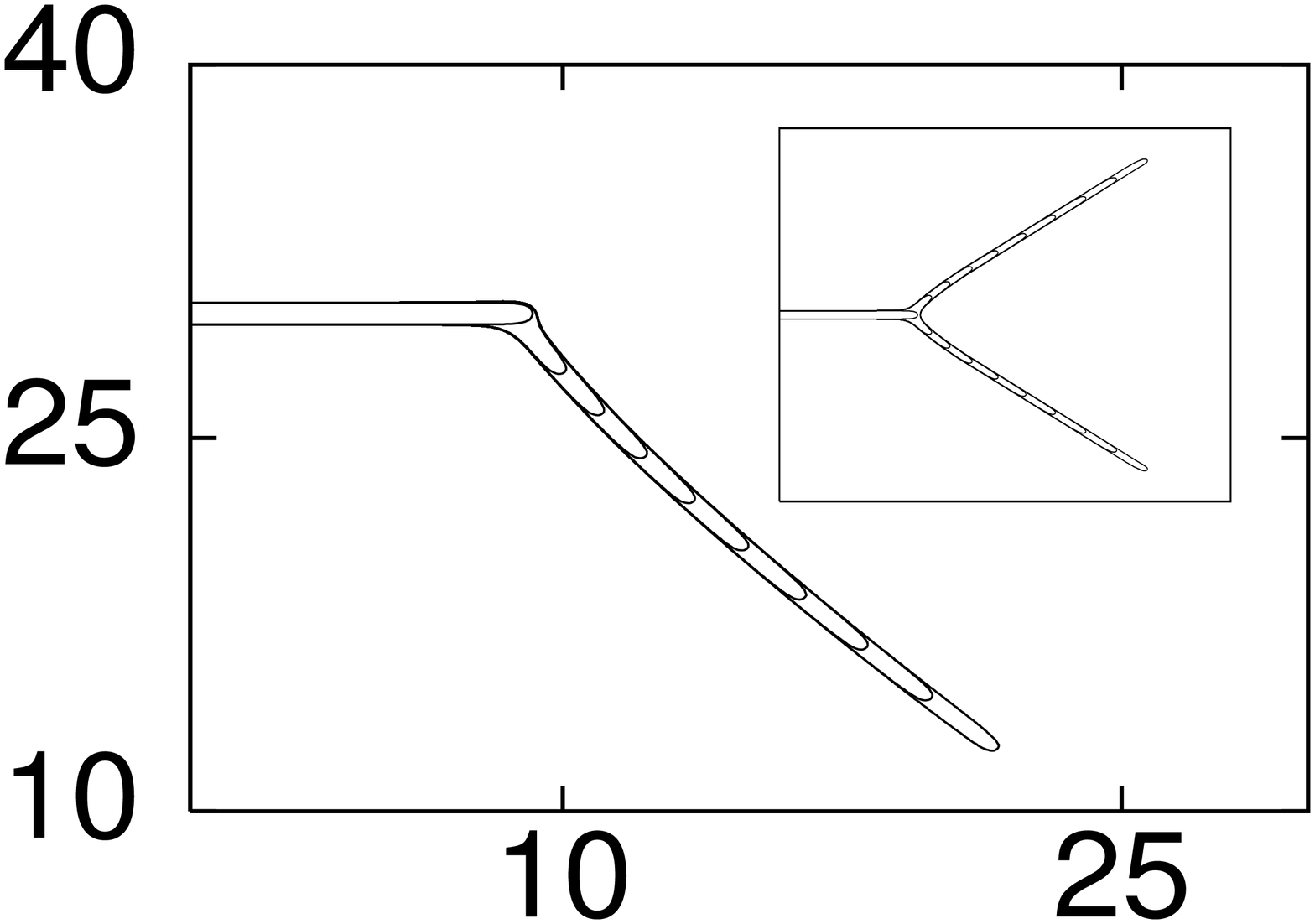}
\caption{\label{figbranching}Left: contour plot of the line $\phi=0.5$
taken at 1. time unit (tu) intervals during the branching of a crack under
mode I loading with $\Delta_y=18.$ and $\beta=1$. The angle of the two
branches at the branching site is approximately $40^o$ and the critical
speed for which the instability occurs is approximately $0.5$ while the shear
speed is 1, which is in agreement with predictions in
\cite{addabranching}. During the evolution of the system, the lower branch
will recede while the upper branch will propagate and branch irregularly. The simulations are performed using the undamped model.
Right: contour plots of $\phi=0.5$ taken at 40. time units (t.u.) during the
propagation of a damped crack in a medium where pure mode II loading is
applied. The initial crack is straight and oriented along the $x$ axis.
The crack propagates in a direction that nullifies the mode II stress
intensity factor. Inset: same parameters with
$\alpha=0$. One can see, that in this case, the crack, branches and
propagates in two directions symmetrical with respect to the $x$ axis. The
second branch propagates in a region where the material is strongly
compressed. Simulations are performed using the overdamped model.}
\end{center}
\vspace*{-.45cm}
\end{figure}

\begin{figure}
\includegraphics[width=0.4\textwidth]{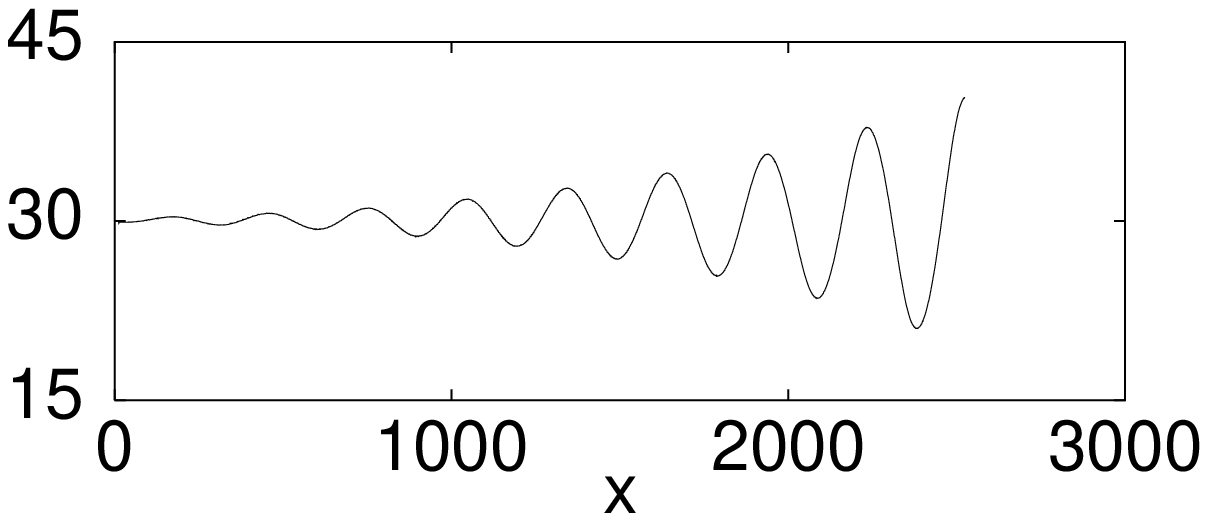}
\includegraphics[width=0.4\textwidth]{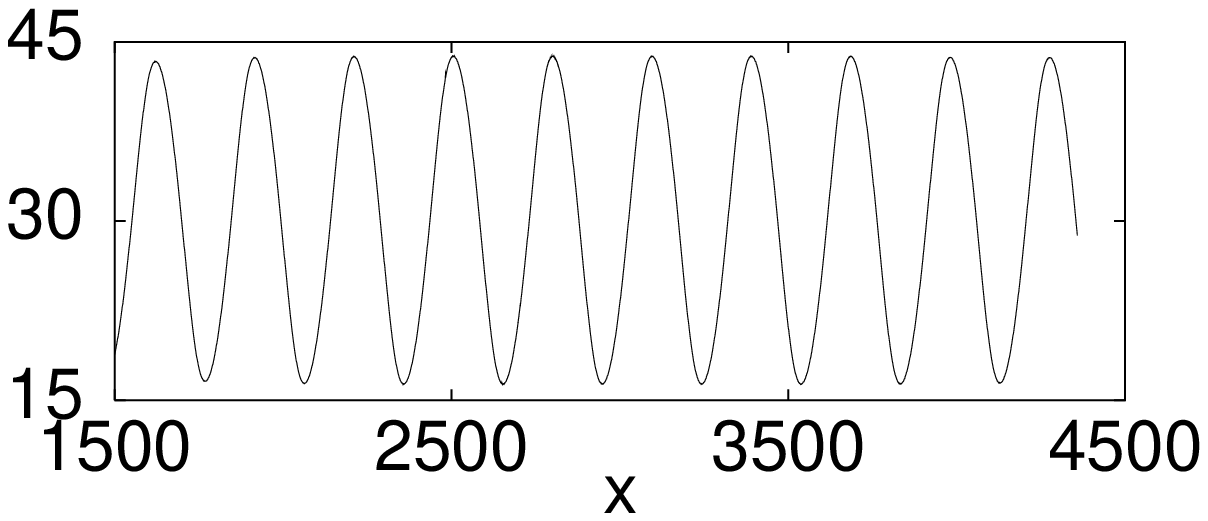}
\includegraphics[width=0.4\textwidth]{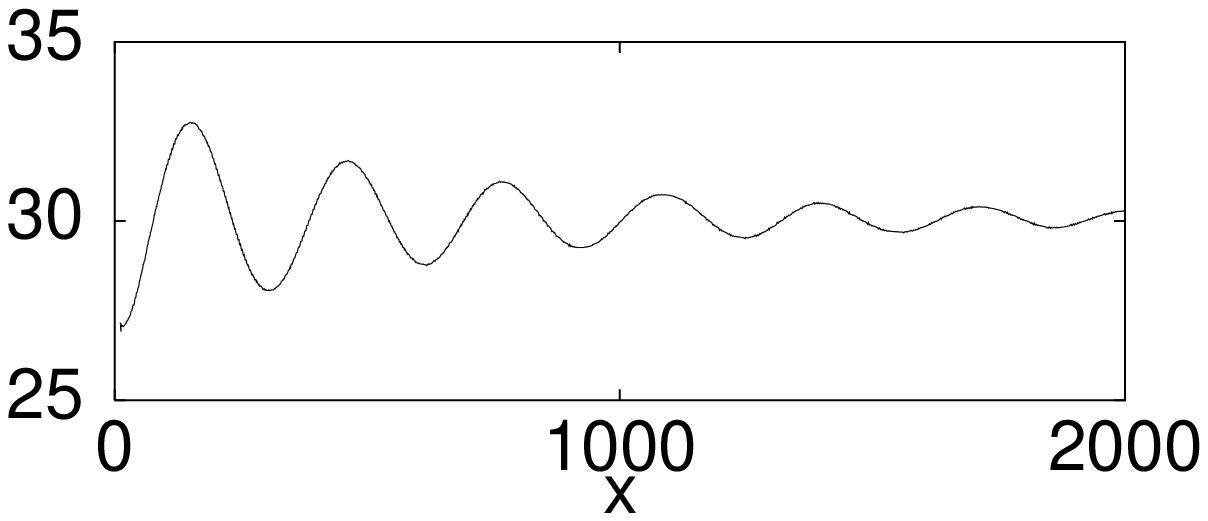}
\caption{\label{tip_traj}Fracture tip trajectories for $\Delta_y=10$ and $r= W\Delta_x/L=8.0$ during the transient regime (a) and when a stationary state is reached (b). The steady-state tip trajectory matches a simple sinusoid. (c) Tip trajectory  for $\Delta_y=10.$ and $r=W\Delta_x/L=7.25$, exhibiting/ damped oscillations. In (b) the mean tip speed is 0.362 and the mean  tip speed along the x axis is 0.354. In (c), the tip speed is 0.347 once straight propagation has resumed. The transverse and longitudinal speeds of sound are respectively 1. and 1.22. The period of the oscillations in (a) is about 831 t.u and the corresponding wavelength is 294 s.u }
\vspace*{-.45cm}
\end{figure}

We first tested the model for damped dynamics obtained by replacing
$\partial _{tt}$ by $\partial _t$.  For pure mode I loading, the fracture
began to propagate once the imposed elastic energy was higher than the
fracture energy (see later).  In addition, in this case, the
stress intensity factor at the tip of a steady crack was found to be
constant for various loading configuration, hence obeying the Irwin
criterion\cite{Freund} (data not shown). More interestingly, numerical
simulations in the case of pure mode II loading showed that the model
respects the local symmetry principle: the fracture propagates in the
direction which nullifies the stress intensity factor for mode II (see
Fig. 2b). Hence this model reproduces well the behavior of a single crack
in the damped regime. Note that this result depends on our asymmetry
parameter $\alpha$; allowing breakage under compression leads to a model
which does not follow the local symmetry principle (Fig. 2b, inset). The results for different $\alpha$ ranging from $1$ to $2$ were similar.

Next, we turn briefly to results obtained in the dynamic (non-damped) case
under pure mode I loading. As expected by analogy with the results of
\cite{Karmafrac} for the case of mode III loading, the initiation of crack
propagation appeared at the Griffith threshold with good accuracy. Indeed,
according to the Griffith criterion, one would expect a crack to begin to
propagate for a value of the $y$ extension $\Delta_y$ bigger than
$\Delta_c$ which is solution of:
\begin{equation}
W\left((\frac{1}{2}\lambda+\mu)\frac{\Delta_c^2}{W^2}\right)=2\int_0^1 d\phi\ \sqrt{2(1-g(\phi))+V(\phi)}
\end{equation}
This formula was derived in \cite{Karmafrac} and follows from the
asymptotic solution of the model in the region far behind the crack tip.
For the parameter values used here, this threshold is at $9.5$ whereas in
our simulations the crack begins to propagate for $\Delta_y$ bigger than
$9.7 \pm 0.1$; this 2\% discrepancy is due to discretization and finite
width effects and gives some measure of the accuracy of our computations.
The speed of the stable crack behaved qualitatively as expected when
loading was increased. When loading was further increased, the dynamic
fracture exhibits a branching instability and a secondary crack begins to
propagate (see Fig \ref{figbranching}). This branching instability is
compatible with what has been observed experimentally in a wide range of
materials \cite{branch1,branch2}. A complete analysis of the branching
phenomenon will be addressed in further work.

We now turn to our major interest here, the case of a dynamic crack
propagating under biaxial stress. Experimental work by Deegan \textit{et
al.}\cite{Swinney2002} has shown that for a given imposed strain in the
$y$ direction (see Fig. \ref{shemaswin}), there is a threshold value of
the $x$ strain for which the crack propagation is no longer straight;
instead, the crack tip position begins to oscillate. In fact, the
instability appears to be supercritical and the tip trajectory is
well-approximated by a sinusoidal line with finite wavelength and
amplitude.  Recall that in our calculations strains are applied by moving
the rightmost border of the sheet by $\Delta_x$ and by moving the top
border by $\Delta_y$; hence, if $\Delta_x=0$, the system is set to pure
mode I.  The experimental results translate into the prediction of a Hopf
bifurcation that should occur as we cross a threshold value of
$r=W\Delta_x/L$, with $\Delta_y$ being fixed.
\begin{figure}
\includegraphics[width=0.4\textwidth,height=0.15\textwidth]{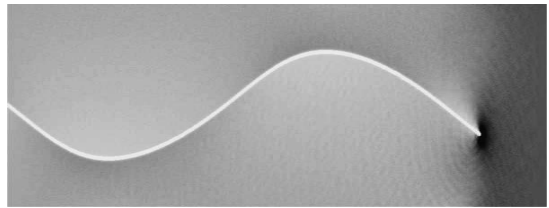}
\caption{Elastic energy landscape during the propagation of an oscillating crack. The white region corresponds to the broken region ($\phi<0.5$) and the gray scale represents the density of elastic energy with dark regions corresponding to high elastic energy density. Parameter values are as in figure \ref{tip_traj} (a). Between the arches of the sinusoidal line of the crack, the elastic energy density is lower than in the vicinity of the upper and lower boundaries of the slice.}
\vspace*{-.45cm}
\end{figure}

The results of our numerical simulations for two sets of $\lambda,\ \mu$
and different values of $\Delta_y$ faithfully reproduces the aforementioned
phenomenology. Namely, the fracture tip trajectory indeed undergoes a Hopf
a bifurcation when the $x$ extension is increased over a threshold value
that depends on both parameter regime and the vertical extension. This
bifurcation is characterized by the fact that below threshold, the tip
position shows damped oscillations (see figure \ref{tip_traj}) and ends up
propagating along a straight line, whereas above threshold those
oscillations are amplified and the restabilized state corresponds to the
situation where the fracture tip oscillates at a finite wavelength with a
finite amplitude (see figure \ref{tip_traj}). We checked that this
instability was not due to waves reflecting at the boundaries that can
create periodic markings called Wallner lines\cite{Lawn}. Indeed, the
expected wavelength of such markings would be
$\lambda=Wv/c\sqrt{1-v^2/c^2}$, that is about 10 s.u. while the wavelength
observed here is about 300 s.u. This is confirmed by the fact that
switching to quasi-absorbing boundary conditions at the $y=W$ and $y=0$
lines does not affect the oscillations.

When the restabilized state is reached, the trajectory of the tip is
almost indistinguishable from a sinusoidal line, as in \cite{Swinney2002}.
One can also note that the horizontal tip speed oscillates with a
frequency equal to two times the frequency of the vertical position, so
that the maximum of the horizontal tip speed are reached when the
instantaneous tip velocity is directed along the $x$ axis. In addition, the tip speed tangent to its trajectory is kept almost constant (up to numerical errors) and for different values of $r$ (r varying between 7 and 8), we did not find significant changes in the tip speed (less than a few \%). A picture of this state is presented in Fig. 4.

\begin{figure}
\includegraphics[width=0.23\textwidth]{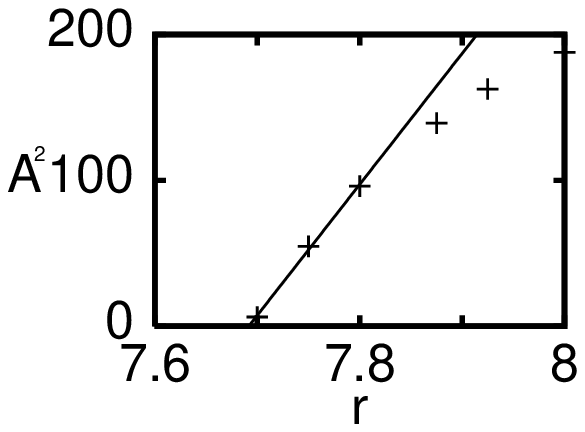}
\includegraphics[width=0.23\textwidth]{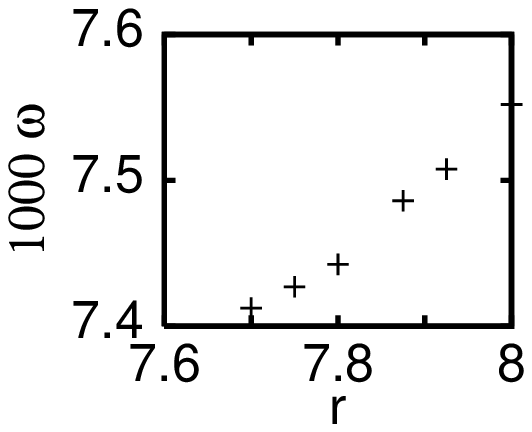}
\begin{center}
\caption{\label{evstrainx} (a): Square of the amplitude of the oscillations 
as a function of $r=W \Delta_x/L$. (b): frequency $\omega$ (+)   of the oscillations. The wavelength (not shown) varies like the inverse of $\omega$ since the crack speed is not varying much (about $10^{-3}$) over the parameter range presented here.}
\end{center}
\vspace*{-.45cm}
\end{figure}

We now describe the changes in the oscillating restabilized state when the
strain along the $x$ axis is increased. As seen in figure \ref{evstrainx},
the amplitude of the oscillations behaves like
$\sqrt{\Delta_x-\Delta_{xc}}$ close to threshold, which is consistent with 
 a supercritical Hopf bifurcation, as
experimentally observed in \cite{Swinney2002}. The wavelength and period
of the tip oscillation decreases slightly when ${\Delta_x-\Delta_{xc}}$ is
increased; this differs from results in \cite{Swinney2002} where the
wavelength increases when ${\Delta_x-\Delta_{xc}}$. This may be due to
non-linear elasticity effects present in the experiment.

We also performed numerical simulations with biaxial strain and
$\alpha=0$.The results obtained did not differ significantly from
that observed for $\alpha=1.5$. This is somewhat surprising, since the simplest interpretation of
the oscillations observed here suggests that the mechanism is at least partially similar to the one underlying oscillations of a quasi-static crack propagating
in a thermal gradient\cite{pommeauadda,naturegradient}. In that case,
theoretical work has explained the transition to an oscillating crack using a
(modified) principle of local symmetry\cite{sethnapls,osci2003}; changing the tip direction induces a mode II component which causes further deviation from the original line. We have already shown that this principle does not apply for a symmetric model for pure mode-II loading. Perhaps the explicit breaking of the symmetry by the mode-I part of the driving is enough to suppress the unphysical compressional breaking for the case of $\alpha=0$, and hence this model still exhibits the Hopf bifurcation. In support of this, we verified that even in this case,  a crack tip with damped dynamics will obey the principle of local symmetry, if in addition to a pure mode II load one
adds a small mode I extension. We should note, though, that we never observe oscillations with damped dynamics, even when $r$ was set to very close to $\Delta_y$, i. e. close to hydrostatic strain. However increasing $\tau$ (up to 50), i.e. increasing the dissipation at the crack tip did not affect significantly either the instability wavelength of the oscillating crack or the threshold but did reduce the crack speed (from 0.362 to 0.05 ($\tau=50$)).  Also, changes in $\Delta_y$ ($\Delta_y=12, 14$, with $\tau=20$ to avoid branching) did not affect  significantly the wavelength of the oscillating crack but did change the threshold. Interestingly, the wavelength scales linearly with  $W$.  Hence, it seems that the saturation of the amplitude is governed by the interaction of the tip with the sidewall.  Underlying the actual instability is perhaps the simple fact that an oscillating crack will alleviate extra elastic energy under biaxial strain.

In summary, this paper shows that the extension of the KKL phase field
model of crack propagation to full vector elasticity qualitatively
reproduces the different instabilities observed when considering the
propagation of cracks. One should note that with an extremely simple model
based on generic physical considerations we were nonetheless able to
reproduce the variety of observed patterns. This lends us confidence in
the entire modeling approach and suggests that we proceed in two
complementary directions. First, we can continue to investigate the
phenomenology of KKL, specifically looking at the interaction of different
cracks
 (the phase field method can easily deal with intersecting interfaces) and also truly three-dimensional effects\cite{nature3d}. At the same time, it is time to begin understanding how to combine this method with microscopic interaction data about specific materials so as to enable the building of more quantitatively reliable models of dynamic fracture.

It is a pleasure to acknowledge useful discussions with A. Karma and D. Kessler.
This research is supported by the National Science Foundation through
Grant No. DMR-0101793.


\begin{thebibliography}{17}
\expandafter\ifx\csname natexlab\endcsname\relax\def\natexlab#1{#1}\fi
\expandafter\ifx\csname bibnamefont\endcsname\relax
  \def\bibnamefont#1{#1}\fi
\expandafter\ifx\csname bibfnamefont\endcsname\relax
  \def\bibfnamefont#1{#1}\fi
\expandafter\ifx\csname citenamefont\endcsname\relax
  \def\citenamefont#1{#1}\fi
\expandafter\ifx\csname url\endcsname\relax
  \def\url#1{\texttt{#1}}\fi
\expandafter\ifx\csname urlprefix\endcsname\relax\def\urlprefix{URL }\fi
\providecommand{\bibinfo}[2]{#2}
\providecommand{\eprint}[2][]{\url{#2}}

\bibitem[{\citenamefont{Fineberg and Marder}(1999)}]{marderreview}
\bibinfo{author}{\bibfnamefont{J.}~\bibnamefont{Fineberg}} \bibnamefont{and}
  \bibinfo{author}{\bibfnamefont{M.}~\bibnamefont{Marder}},
  \bibinfo{journal}{Phys. Rep.} \textbf{\bibinfo{volume}{313}},
  \bibinfo{pages}{2} (\bibinfo{year}{1999}).

\bibitem[{\citenamefont{Kessler and Levine}(2003)}]{KesslerLevine}
\bibinfo{author}{\bibfnamefont{D.}~\bibnamefont{Kessler}} \bibnamefont{and}
  \bibinfo{author}{\bibfnamefont{H.}~\bibnamefont{Levine}},
  \bibinfo{journal}{Phys. Rev. E} \textbf{\bibinfo{volume}{68}},
  \bibinfo{pages}{036118} (\bibinfo{year}{2003}).

\bibitem[{\citenamefont{Karma et~al.}(2001)\citenamefont{Karma, Kessler, and
  Levine}}]{Karmafrac}
\bibinfo{author}{\bibfnamefont{A.}~\bibnamefont{Karma}},
  \bibinfo{author}{\bibfnamefont{D.}~\bibnamefont{Kessler}}, \bibnamefont{and}
  \bibinfo{author}{\bibfnamefont{H.}~\bibnamefont{Levine}},
  \bibinfo{journal}{Phys. Rev. Lett.} \textbf{\bibinfo{volume}{87}},
  \bibinfo{pages}{045501} (\bibinfo{year}{2001}).

\bibitem[{\citenamefont{Karma and Lobkovski}()}]{Karmabranching}
\bibinfo{author}{\bibfnamefont{A.}~\bibnamefont{Karma}} \bibnamefont{and}
  \bibinfo{author}{\bibfnamefont{A.}~\bibnamefont{Lobkovski}},
  \bibinfo{note}{\textit{Unsteady crack motion and branching in a phase field
  model} cond-mat/0401056}.

\bibitem[{\citenamefont{Deegan et~al.}(2002)\citenamefont{Deegan, Petersan,
  Marder, and Swinney}}]{Swinney2002}
\bibinfo{author}{\bibfnamefont{R.}~\bibnamefont{Deegan}},
  \bibinfo{author}{\bibfnamefont{P.}~\bibnamefont{Petersan}},
  \bibinfo{author}{\bibfnamefont{M.}~\bibnamefont{Marder}}, \bibnamefont{and}
  \bibinfo{author}{\bibfnamefont{H.}~\bibnamefont{Swinney}},
  \bibinfo{journal}{Phys. Rev. Lett.} \textbf{\bibinfo{volume}{88}},
  \bibinfo{pages}{014304} (\bibinfo{year}{2002}).

\bibitem[{\citenamefont{Aranson et~al.}(2000)\citenamefont{Aranson, Kalatsky,
  and Vonokur}}]{igorphasefield}
\bibinfo{author}{\bibfnamefont{I.~S.} \bibnamefont{Aranson}},
  \bibinfo{author}{\bibfnamefont{V.~A.} \bibnamefont{Kalatsky}},
  \bibnamefont{and} \bibinfo{author}{\bibfnamefont{V.~M.}
  \bibnamefont{Vonokur}}, \bibinfo{journal}{Phys. Rev. Lett.}
  \textbf{\bibinfo{volume}{85}}, \bibinfo{pages}{118} (\bibinfo{year}{2000}).

\bibitem[{\citenamefont{Eastgate et~al.}(2002)\citenamefont{Eastgate, Sethna,
  Rauscher, Cretegny, Chen, and Myers}}]{sethnaphasefield}
\bibinfo{author}{\bibfnamefont{L.}~\bibnamefont{Eastgate}},
  \bibinfo{author}{\bibfnamefont{J.}~\bibnamefont{Sethna}},
  \bibinfo{author}{\bibfnamefont{M.}~\bibnamefont{Rauscher}},
  \bibinfo{author}{\bibfnamefont{T.}~\bibnamefont{Cretegny}},
  \bibinfo{author}{\bibfnamefont{C.-S.} \bibnamefont{Chen}}, \bibnamefont{and}
  \bibinfo{author}{\bibfnamefont{C.}~\bibnamefont{Myers}},
  \bibinfo{journal}{Phys. Rev. E} \textbf{\bibinfo{volume}{65}},
  \bibinfo{pages}{36117} (\bibinfo{year}{2002}).

\bibitem[{\citenamefont{Adda-Bedia}()}]{addabranching}
\bibinfo{author}{\bibfnamefont{M.}~\bibnamefont{Adda-Bedia}},
  \bibinfo{note}{\textit{Brittle fracture dynamics with arbitrary paths : III.
  The branching instability under general loading} (2003)}.

\bibitem[{\citenamefont{Freund}(1990)}]{Freund}
\bibinfo{author}{\bibfnamefont{L.}~\bibnamefont{Freund}},
  \emph{\bibinfo{title}{Dynamic Fracture Mechanics}}
  (\bibinfo{publisher}{Cambridge University Press (UK)}, \bibinfo{year}{1990}).

\bibitem[{\citenamefont{Fineberg et~al.}(1991)\citenamefont{Fineberg, Gross,
  Marder, and Swinney}}]{branch1}
\bibinfo{author}{\bibfnamefont{J.}~\bibnamefont{Fineberg}},
  \bibinfo{author}{\bibfnamefont{S.~P.} \bibnamefont{Gross}},
  \bibinfo{author}{\bibfnamefont{M.}~\bibnamefont{Marder}}, \bibnamefont{and}
  \bibinfo{author}{\bibfnamefont{H.~L.} \bibnamefont{Swinney}},
  \bibinfo{journal}{Phys. Rev. Lett.} \textbf{\bibinfo{volume}{67}},
  \bibinfo{pages}{457} (\bibinfo{year}{1991}).

\bibitem[{\citenamefont{Ravi-Chandar and Knauss}(1984)}]{branch2}
\bibinfo{author}{\bibfnamefont{K.}~\bibnamefont{Ravi-Chandar}}
  \bibnamefont{and} \bibinfo{author}{\bibfnamefont{W.~G.}
  \bibnamefont{Knauss}}, \bibinfo{journal}{Int. J. Fracture}
  \textbf{\bibinfo{volume}{25}}, \bibinfo{pages}{247} (\bibinfo{year}{1984}).

\bibitem[{\citenamefont{Lawn}(1993)}]{Lawn}
\bibinfo{author}{\bibfnamefont{B.}~\bibnamefont{Lawn}},
  \emph{\bibinfo{title}{Fracture of Brittle solid, second edition}}
  (\bibinfo{publisher}{Cambridge University Press (UK)}, \bibinfo{year}{1993}).

\bibitem[{\citenamefont{Adda-Bedia and Pomeau}(1995)}]{pommeauadda}
\bibinfo{author}{\bibfnamefont{M.}~\bibnamefont{Adda-Bedia}} \bibnamefont{and}
  \bibinfo{author}{\bibfnamefont{Y.}~\bibnamefont{Pomeau}},
  \bibinfo{journal}{Phys. Rev. E} \textbf{\bibinfo{volume}{52}},
  \bibinfo{pages}{4105} (\bibinfo{year}{1995}).

\bibitem[{\citenamefont{Yuse and Sano}(1993)}]{naturegradient}
\bibinfo{author}{\bibfnamefont{A.}~\bibnamefont{Yuse}} \bibnamefont{and}
  \bibinfo{author}{\bibfnamefont{M.}~\bibnamefont{Sano}},
  \bibinfo{journal}{Nature} \textbf{\bibinfo{volume}{362}},
  \bibinfo{pages}{329} (\bibinfo{year}{1993}).

\bibitem[{\citenamefont{Hodgdon and Sethna}(1993)}]{sethnapls}
\bibinfo{author}{\bibfnamefont{J.}~\bibnamefont{Hodgdon}} \bibnamefont{and}
  \bibinfo{author}{\bibfnamefont{J.}~\bibnamefont{Sethna}},
  \bibinfo{journal}{Phys. Rev. B} \textbf{\bibinfo{volume}{47}},
  \bibinfo{pages}{4831} (\bibinfo{year}{1993}).

\bibitem[{\citenamefont{Bouchbinder et~al.}(2003)\citenamefont{Bouchbinder,
  Hentschel, and Procaccia}}]{osci2003}
\bibinfo{author}{\bibfnamefont{E.}~\bibnamefont{Bouchbinder}},
  \bibinfo{author}{\bibfnamefont{H.}~\bibnamefont{Hentschel}},
  \bibnamefont{and}
  \bibinfo{author}{\bibfnamefont{I.}~\bibnamefont{Procaccia}},
  \bibinfo{journal}{Phys. Rev. E} \textbf{\bibinfo{volume}{68}},
  \bibinfo{pages}{036601} (\bibinfo{year}{2003}).

\bibitem[{\citenamefont{Sharon et~al.}(2001)\citenamefont{Sharon, Cohen, and
  Fineberg}}]{nature3d}
\bibinfo{author}{\bibfnamefont{E.}~\bibnamefont{Sharon}},
  \bibinfo{author}{\bibfnamefont{G.}~\bibnamefont{Cohen}}, \bibnamefont{and}
  \bibinfo{author}{\bibfnamefont{J.}~\bibnamefont{Fineberg}},
  \bibinfo{journal}{Nature} \textbf{\bibinfo{volume}{410}}, \bibinfo{pages}{68}
  (\bibinfo{year}{2001}).

\end{thebibliography}
\end{document}